# The Single Differential Cross Sections (SDCS) for H(3s) Ionization in the First-Born Approximation by Electron and Positron Impact

[1]Fahadul Islam, [2] Sunil Dhar

Department of Mathematics, Chittagong University of Engineering and Technology, Chittagong, Bangladesh
Email: [1]fahadulislambgd@gmail.com, [2]sdhar@cuet.ac.bd

**Abstract:** A theoretical study was conducted on the impact of electron and positron impact ionization of excited hydrogen atoms that were in the 3s state; this study was conducted within the First-Born Approximation (FBA), which provides an analytical expression for the transition matrix in terms of the Bethe-Lewis Integral Formalism. This formalism utilized both Coulomb continuum and confluent hypergeometric functions to describe the scattering states involved. Single Differential Cross Sections (SDCS) were calculated for incident energies of 100, 150, 200, and 250 eV. The data obtained indicated a peak in the ionization rates approximately at 200 eV, with the ionization rate decreasing as the incident energy increased further. The diffuse radial nature of the 3s wave function is shown to increase the sensitivity of the ionization dynamics to the incident particle energy. Asymmetries in charge were also detected; specifically, at low energy of the ejected electron, the SDCS values for positrons were greater than the corresponding values for electrons; however, as the energy of the incident particles was increased, these differences disappeared, thereby demonstrating the applicability of the FBA at high energy limits. The residual differences at low energy were due to the omission of exchange and post-collision interactions from the model. The results of this work can be used as benchmarking for the development of more complex distorted wave and multi-scattering theories in excited state ionization processes.

## 1. Introduction

The ionization of atomic hydrogen by charged-particle impact is among the simplest problems of atomic collision theory and provides a criterion for the test of the accuracy of quantum mechanics scattering modes [1-3]. The simplicity of the system, because of its structure of only one electron, permits a comparison between the experiment and theory in the way of accuracy and also gives some knowledge of the more fundamental processes of ionization dynamics [4], [5]. Understanding the mechanisms of these processes is extremely significant not only in physics but also in astrophysics, as well as in studying the physical processes involved in controlled fusion devices and antimatter interactions with matter [6], [7].

The theoretical basis for the treatment of the ionization by charged particles was provided by Bethe [1], who, by his pioneering development of quantum mechanics and of high-energy scattering, laid the foundation for perturbation theory of the modern type. This theory was later advanced by Lewis [2] in the second Born approximation so as to have the effect of the potential scattering at the intermediate energies of the discharged particles and thus get the predictive power. This setup served as a basis for the work of Das and Seal [3], [4], who later treated the multiple scattering and the first-born approximation (FBA) methods, and for their treatment of the electron-hydrogen ionization in a systematic and thorough manner at intermediate and high energies. These results provided the basis for further theoretical research on the subject of single (SDCS) and double differential cross-sections (DDCS). Experimental work has been equally as active as theoretical work, and the results of this work provide necessary data for testing the models. Shyn has engaged in extensive investigations [5] regarding the double differential cross-sections for hydrogen and molecular targets in the 25 eV to 400 eV range and has arrived at certain empirical laws to be referred to again. The integrated cross-section analysis carried out by Konovalov and McCarthy has shown a more optimistic view compared with theoretical predictions. The integrated cross-sections of these experiments have verified the accuracy of the Born-type approximations, particularly in the intermediate- to high-energy range.

The investigation carried out by Das and Dhar has presented the necessary statistical development of the theoretical data regarding metastable hydrogen-type states (2s, 2p) and decomposed excited states [8-10]. The first-born theory, with symmetrical geometrical arrangements, has been successfully used in the observations concerning ionization of the K-shell and various states of excitation. The communications of Vucic et al. [11] and Hafid et al. [12] have presented new second Born-type and correlated theories of (e, 2e)-collisions, which show considerable changes in the ionization details of the target atom, which vary with the states of excitation and the geometry existing.

In the last few years excited hydrogen atoms have drawn increased attention, and particularly those in n = 2 type shells, because of their great importance in connection with the plasma phenomenon and high-energy processes in general, which take place in the various astronomical environments [13-17]. They have been studied systematically by Dhar and Nahar [18], [19], et al., with respect to the excitation of the states 2p, 3p, and [17] 3p and 3d by means of the First Born and multiple scattering methods, in which the treatment of the exchange and post-collision effects receives the deserved attention. Attention has been directed also upon the excitation of the state [3-8] by Banerjee et al. [20] and Islam et al. [21], advanced methods of computation in the generation of double differential cross-section data, which approximate in soundness of error nearly to the existing theoretical. The hydrogen atom in the [3-8] state, therefore, is peculiarly fitted to give a study of the ionization phenomena, which are sensitive to the excitation conditions and characters. The fact that the diffuse character of the radial wave function is a minimum, and that the degree of binding is of a low order, precludes the possibility of this being otherwise; this reacts then to changes in the order of the scattering potentials. Precludes also that the long-range correlations obtained are favored, and in this connection the studies receive a specific difference from all the ground state or lower excited configurations. So far as the hydrogen (3s) atom is concerned comparatively little theoretical work has been expended on regarding the single differential cross sections (SDCS.) in the case of ionization by means of projectile electrons and positrons respectively, and experiments carried out in the latter instance will be found to assume potentials of and will also require importance of this charge sign existent due to the assumed symmetry and asymmetry relatively existent due to the attractive and repulsive character of the Coulomb force existing between the bombarding projectile and the atom of the target, and moreover will be also of value from the point of view of testing the first object of the application of perturbated theory at moderate energy [7], [14, 15].

In the present paper it is proposed to consider the question of SDCS. for either electron or positron impact on the hydrogen (3s), allowing a considerable amount of detail of a theoretical character in regard to the application of the First-Born Approximation. The analytic results can be obtained also by the method of the integral of Bethe and Lewis. Methods will be improved on such a form as to give the arguments directly for the corresponding in other forms and conditions of strength with the final states of the various physics referred to under the respective headings of wave function. The latter can be obtained either from the method of the continuum-Coulomb or of the confluent hypergeometric series of functions. The numerical valuations will be made for the incidents of energies 100 eV, 150 eV, 200 eV, and 250 eV, so giving a wide passage of energies here, which will be found to possess a place in the results gained, undeviating one comparatively independently of any theoretical aspect of treatment or of the experimental kind. The reference to the effect will be seen to arise by comparing the SDCS. so far accepted in the more established, accurate, derived, integrated reconnaissance results and would be found to react satisfactorily insofar as theoretical results are concerned regarding the derivatives as given in previous papers of ours [3-6], [8-10], [18-21], so disclosing some new side of the matter regarding the excitation dependence regarding the projectile charge sign asymmetry, so producing considerations as to the essential dynamics involved in the collision of impact of electrons and positrons together with ionization of atomic hydrogen.

## 2. Theory

### 2.1 Theoretical Framework

The ionization of metastable 3s excited hydrogen atoms by electron and positron impact is considered in the quantum mechanical First-Born Approximation (FBA). In this method, the projectile interacts only once with the target, thereby causing a single-step perturbation that makes the problem both mathematically tractable and physically clears [1-3].

The ionization processes may be represented as:

$e^- + H(3s) \to H^+ + 2e^-$  (1)

$e^+ + H(3s) \to H^+ + 2e^-$  (2)

Here 3s denotes the metastable state of hydrogen. The outgoing products are the correlated three-body Coulomb continuum. This single interaction assumption is valid for incident energies above about 100 eV, where higher-order multiple integrations are weak by comparison [4–6].

### 2.2 First-Order Transition Amplitude

In the FBA the transition amplitude, called the T-matrix element, is defined by the overlap of the unperturbed initial and final channel states through the interaction potential:

$T_{FI} = \langle \Psi_F^{(-)}(\bar{\gamma}_a, \bar{\gamma}_b) | V_I(\bar{\gamma}_a, \bar{\gamma}_b) | \Phi_I(\bar{\gamma}_a, \bar{\gamma}_b) \rangle$  (3)

The incident state $\Phi_I(\bar{\gamma}_a, \bar{\gamma}_b)$ is a plane wave for the incident projectile, and the bound hydrogenic orbital $\phi_{3s}(\bar{\gamma}_a)$, the final state $\Psi_F^{(-)}(\bar{\gamma}_a, \bar{\gamma}_b)$, depicts two continuum particles, either 2e's or an electron-positron pair, distorted by the long-range Coulomb potential. The interaction potential $V_I(\bar{\gamma}_a, \bar{\gamma}_b)$ comprises the projectile-electron and projectile-nuclear s states:

$V_I(\bar{\gamma}_a, \bar{\gamma}_b) = \frac{Z}{\gamma_b} - \frac{Z}{\gamma_{ab}}$  (4)

Where "Z= -1 for electrons" is the nuclear charge of the hydrogen atom, $\bar{\gamma}_a$ and $\bar{\gamma}_b$ are the distances of the two electrons from the nucleus, and $\gamma_{ab}$ is the distance between the two electrons.

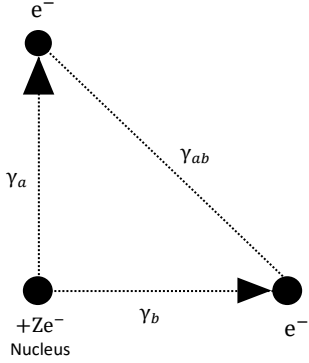

**Figure a**: Collision effect amongst two electrons and the nucleus.

### 2.3 Bound and Continuum Wave functions

The initial channel unperturbed wave function is

$$\Phi_I(\bar{\gamma}_a, \bar{\gamma}_b) = \frac{e^{i.\bar{p}_i.\bar{\gamma}_b}}{(2\pi)^{\frac{3}{2}}} \phi_{3s}(\bar{\gamma}_a)$$

$$= \frac{e^{i.\bar{p}_i.\bar{\gamma}_b}}{(2\pi)^{\frac{3}{2}}} \cdot \frac{1}{81\sqrt{3\pi}} (27 - 18\gamma_a + 2\gamma_a^2) e^{-\lambda_a \gamma_a} \quad (5)$$

Here,

$$\phi_{3s}(\bar{\gamma}_a) = \frac{1}{81\sqrt{3\pi}} (27 - 18\gamma_a + 2\gamma_a^2) e^{-\lambda_a \gamma_a} \quad (6)$$

Here $\lambda_a = \frac{1}{3}$, $\phi_{3s}(\bar{\gamma}_a)$ is the hydrogen 3s-state wave function, and $\Psi_F^{(-)}(\bar{\gamma}_a, \bar{\gamma}_b)$ is the final three-particle scattering state wave function [03] with the electrons being in the continuum with momenta $\bar{p}_a$ and $\bar{p}_b$. And the coordinates of the two electrons are $\bar{\gamma}_a$ and $\bar{\gamma}_b$ respectively. Here the approximate wave function $\Psi_F^{(-)}(\bar{\gamma}_a, \bar{\gamma}_b)$ is given by

$$\Psi_F^{(-)}(\bar{\gamma}_a, \bar{\gamma}_b) =$$
$$N(\bar{p}_a, \bar{p}_b)[\phi_{\bar{p}_a}^{(-)}(\bar{\gamma}_a)e^{i\bar{p}_b.\bar{\gamma}_b} + \phi_{\bar{p}_2}^{(-)}(\bar{\gamma}_b)e^{i\bar{p}_a.\bar{\gamma}_a} + \phi_{\bar{p}}^{(-)}(\bar{\gamma})e^{i\bar{P}.\bar{R}} - 2e^{i\bar{p}_a.\bar{\gamma}_a + i\bar{p}_b.\bar{\gamma}_b}]/(2\pi)^3 \quad (7)$$

Here $\bar{\gamma} = \frac{\bar{\gamma}_b - \bar{\gamma}_a}{2}$, $\bar{R} = \frac{\bar{\gamma}_b + \bar{\gamma}_a}{2}$, $\bar{p} = (\bar{p}_b - \bar{p}_a)$, $\bar{P} = (\bar{p}_b + \bar{p}_a)$

Here $N(\bar{p}_a, \bar{p}_b)$ is the normalization constant, given by,

$$|N(\bar{p}_a, \bar{p}_b)|^{-2} = \left| 7 - 2[\lambda_a + \lambda_b + \lambda_c] - \left[\frac{2}{\lambda_a} + \frac{2}{\lambda_b} + \frac{2}{\lambda_c}\right] + \left[\frac{\lambda_a}{\lambda_b} + \frac{\lambda_a}{\lambda_c} + \frac{\lambda_b}{\lambda_a} + \frac{\lambda_b}{\lambda_c} + \frac{\lambda_c}{\lambda_a} + \frac{\lambda_c}{\lambda_b}\right] \right| \quad (8)$$

Here,

$$\lambda_a = e^{\frac{\pi\alpha_a}{2}} \Gamma(1 - i\alpha_a), \quad \alpha_a = \frac{1}{p_a}$$
$$\lambda_b = e^{\frac{\pi\alpha_b}{2}} \Gamma(1 - i\alpha_b), \quad \alpha_b = \frac{1}{p_b}$$
$$\lambda_c = e^{\frac{\pi\alpha}{2}} \Gamma(1 - i\alpha), \quad \alpha = -\frac{1}{p}$$

Here $\phi_{\bar{q}}^{(-)}(\bar{\gamma})$ is the coulomb wave function, given by, $\phi_{\bar{q}}^{(-)}(\bar{\gamma}) =$

$$e^{\frac{\pi\alpha}{2}} \Gamma(1 + i\alpha) e^{i\bar{q}.\bar{\gamma}} {}_1F_1(-i\alpha, 1, -i[q\gamma + \bar{q}.\bar{\gamma}]) \quad (9)$$

Now applying equations (4), (5), (6) and (7) to the equation (3), we get

$$T_{FI} = N(\bar{p}_a, \bar{p}_b)[T_B + T_{B'} + T_I - 2T_{PB}] \quad (10)$$

Where,

$$T_B = \langle \phi_{\bar{p}_a}^{(-)}(\bar{\gamma}_a) e^{i\bar{p}_b.\bar{\gamma}_b} |V_i| \Phi_i(\bar{\gamma}_a, \bar{\gamma}_b) \rangle \quad (11)$$

$$T_{B'} = \langle \phi_{\bar{p}_b}^{(-)}(\bar{\gamma}_b) e^{i\bar{p}_a.\bar{\gamma}_a} |V_i| \Phi_i(\bar{\gamma}_a, \bar{\gamma}_b) \rangle \quad (12)$$

$$T_I = \langle \phi_p^{(-)}(\bar{\gamma}) e^{i\bar{P}.\bar{R}} |V_i| \Phi_i(\bar{\gamma}_a, \bar{\gamma}_b) \rangle \quad (13)$$

$$T_{PB} = \langle e^{i\bar{p}_a.\bar{\gamma}_a + i\bar{p}_b.\bar{\gamma}_b} |V_i| \Phi_i(\bar{\gamma}_a, \bar{\gamma}_b) \rangle \quad (14)$$

For the first-born approximation, equation (8) may be written as

$$T_B = \frac{1}{162\sqrt{6}\pi^2} \langle \phi_{\bar{p}_a}^{(-)}(\bar{\gamma}_a) e^{i\bar{p}_b.\bar{\gamma}_b} \left| \frac{1}{\gamma_{ab}} - \frac{1}{\gamma_b} \right| e^{i\bar{p}_i.\bar{\gamma}_b} (27 - 18\gamma_a + 2\gamma_a^2) e^{-\lambda_a.\gamma_a} \rangle$$

$$= \frac{1}{162\sqrt{6}\pi^2} \int \phi_{\bar{p}_a}^{(-)*}(\bar{\gamma}_a) e^{i\bar{p}_b.\bar{\gamma}_b} \left| \frac{1}{\gamma_{ab}} - \frac{1}{\gamma_b} \right| e^{i\bar{p}_i.\bar{\gamma}_b} (27 - 18\gamma_a + 2\gamma_a^2) e^{-\lambda_a.\gamma_a} d^3\gamma_a d^3\gamma_b$$

$$T_B = T_{B_1} + T_{B_2} + T_{B_3} + T_{B_4} + T_{B_5} + T_{B_6} \quad (15)$$

Here $T_{B_4} = 0$ and $T_{B_5} = 0$, (for orthogonality condition)

Then putting the values of $T_{B_1}, T_{B_2}, T_{B_3}, T_{B_4}, T_{B_5}$ and $T_{B_6}$ in the equation (15) we get $T_B$.

The direct scattering amplitude $F(\bar{p}_a, \bar{p}_b)$ is then determined from

$$F(\bar{p}_a, \bar{p}_b) = -(2\pi)^2 T_{FI} \quad (16)$$

Similarly for our present study we calculated analytically the above equations using Lewis Integral [2].

### 2.4 Bethe–Lewis Integral Representation

The Bethe-Lewis integral transformation transforms the 6-dimensional T-matrix element into separable radial and angular integrals, thus making the evaluation much simpler [1], [2]. The general Lewis integral is written by,

$${}_1F_1(a, c, z) = \frac{\Gamma(c)}{(a)\Gamma(c-a)} \int_0^1 dx \, x^{(a-1)} (1-x)^{(c-a-1)} e^{(xz)} \quad (17)$$

For the electron impact ionization, the parameters $\alpha_a$, $\alpha_b$ and $\alpha$ are given below,

With $\alpha_a = \frac{1}{P_a}$ for $\bar{q} = \bar{p}_a$, $\alpha_b = \frac{1}{p_b}$ for $\bar{q} = \bar{p}_b$ and $\alpha = -\frac{1}{p}$ for $\bar{q} = \bar{p}$.

For the normalization constant $N(\bar{p}_a, \bar{p}_b)$ of equation (7) has been calculated numerically.

where $_1F_1(a, c, z)$ is the confluent hypergeometric function. The analytic evaluation of given compact forms in terms of logarithmic derivatives of the gamma function, which ensures that stability is retained in the numerical procedure even at the threshold of ionization [9, 10].

**2.5 Cross-Section Formalism**

The Triple Differential Cross Section (TDCS) is related to the T-matrix element by:

$$\frac{d^3\sigma}{dE_a d\Omega_a d\Omega_b} = \frac{p_a p_b}{p_i} |T_{FI}|^2 \quad (18)$$

Where, $E_a$ and $E_b$ are energies and $p_a$, $p_b$ and $p$ are momenta of the outgoing and incoming particles, respectively. The Double Differential Cross Section (DDCS) is obtained by integration of the TDCS over one of the angles of solid emission:

$$\frac{d^2\sigma}{dE_a d\Omega_a} = \int \frac{d^3\sigma}{dE_a d\Omega_a d\Omega_b} d\Omega_b \quad (19)$$

Thus, the Single Differential Cross Section (SDCS) results from the integration of the DDCS over the solid angle $\Omega_a$.

$$\frac{d\sigma}{dE_a} = \int \frac{d^2\sigma}{dE_a d\Omega_a} d\Omega_a \quad (20)$$

These cross-sections represent the differential probabilities of electron (or positron) impact ionization as functions of energy and angle. For the impact of positron ionization, the interchange effects are not manifest since an identifiable molecule of reaction products is produced, while for the electron-impact ionization, they appear through the anti-summarization of the final wavefunction. [11, 12] The different characteristics of Coulomb focusing (electron impact) and attraction post-collision (positive) induce the typical error asymmetries, which are maximum in the area of lesser ejected-electron energies. [13, 16].

**2.6 Numerical Computation**

The previous formal simplifications as a result of the Lewis integral representation show that the necessary radial integrals are obtainable in general in closed form, while the necessary residual angular integrations can be performed numerically. All expressions for TDCS, DDCS, and SDCS have been calculated in the MATLAB computing environment by means of adaptive quadrature, which ensures convergence. All quantities are expressed in atomic units (a.u.).

**2.7 Relation to Earlier Work**

The present procedures represent a generalization of earlier models based on the FBA for metastable and excited hydrogenic states (2s, 2p, 3p, 3d) earlier developed by Das, Dhar, and their co-workers 17−22. The methodology corresponds to the multiple-scattering and coplanar Born-formulation developed by Das & Seal [3, 4] and Dhar & Nahar [18, 19]. Comparison between experimental data 23 of Shyn and theoretical data of Konovalov and McCarthy 24 shows that the model of FBA based on Bethe–Lewis gives predictions for complete cross-sections on an adequate scale both for electron- and positron-impact in processes of ionization of excited H. The fit is very close and endorses that this form has essential validity and, at the same time, serves as the foundation of extensions for higher integrals, relativistic and polarization resolution in the future.

**3. Results And Discussion:**

We used the First-Born Approximation (FBA) to find the Single Differential Cross Sections (SDCS) for ionizing excited hydrogen (3s) with positrons and electrons. We conducted the calculations for incidence energies of 100 eV, 150 eV, 200 eV, and 250 eV, and the results are presented in Figures 1, 2, 3, and 4. Table 1 gives a simplified version of the important numbers. The SDCS are shown for different ejected-electron energies, and an analysis is done to see how the findings change with the ejected electron energy. A contrast is made of how the findings from the electron and positron projectiles are understood.

**3.1 SDCS at 100 eV (Figure 1)**

There is a big peak in the SDCS at the threshold at 100 eV. This is because it is more likely that low-energy electrons will be emitted. This behavior is also seen in Coulomb focusing during electron impact ionization and the post-collision interaction (PCI) effect, when the released electron feels a strong pull from the leftover proton. When the energy of the ejected electron is low relative to the energy of the incoming electron, the SDCS is at its greatest. After that, the SDCS steadily goes down as the energy of the electron that was ejected goes up.

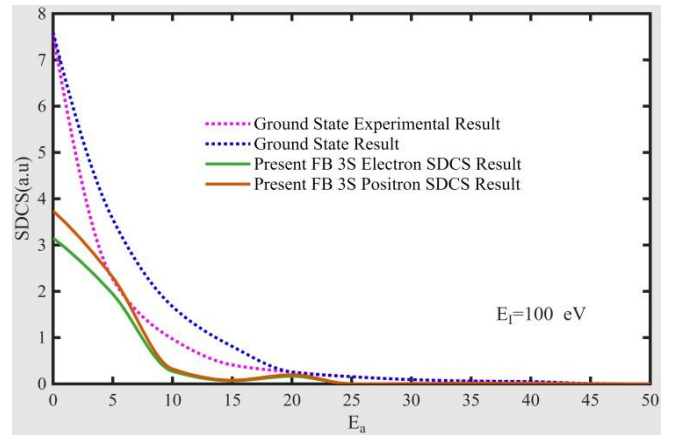

*Figure 1. The First-Born Approximation (FBA) shows the Single Differential Cross Section (SDCS) for the ionization of excited hydrogen (3s) by an electron and positron hitting it with 100 eV of energy. Because of Coulomb focusing and the interaction following the impact, the SDCS exhibits a significant peak at the threshold. The positron impact curve is a little lower than the electron impact curve because the interaction between the projectile and the nucleus pushes them apart. The theoretical patterns are consistent with earlier Born-type and multiple-scattering calculations [3, 5, 14].*

When it comes to positron impact, the SDCS measurements are a little smaller than those for electron impact. This is because the nucleus-positron interaction pushes things away, which makes the threshold cross-section a little less than it is for electrons. The curve caused by the positron, on the other hand, has a greater range at lower energies because the positron and the outgoing electron move around more. This asymmetry qualitatively mirrors that previously identified by Das, Dhar, and Shyn, which is also detected in the metastable ionization of hydrogen, generating a comparable charge-dependent divergence. The overall trend of the result acquired from accurate physics indicates that single-step ionization processes are more prevalent. This kind of behavior is what makes the FBA useful in the intermediate energy range.

### 3.2 SDCS at 150 eV (Figure 2)

The SDCS curves change a lot in shape and size at 150 eV. The electron impact SDCS exhibits a minor reduction in the near-threshold peak and a slower decay. This suggests that the shift from Coulomb to kinetic relation is happening slowly. We can also see that the spectrum of the released electrons is a little off-center. This phenomenon could be because the scattered projectile has higher kinetic energy, which could mean a long, high-energy tail. Using the First-Born Approximation,

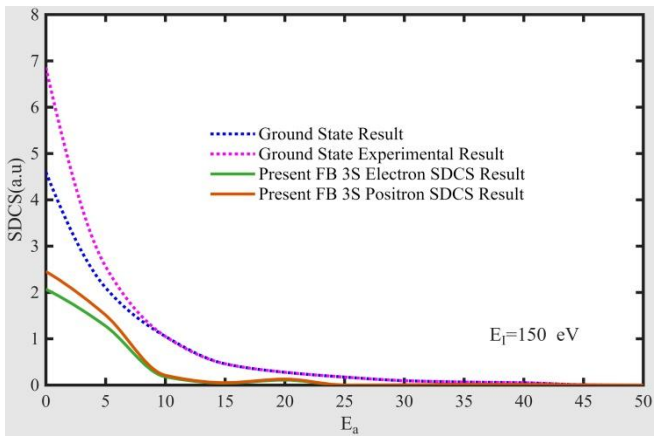

*Figure 2 shows the Single Differential Cross Section (SDCS) for hydrogen (3s) ionization by electrons and positrons at incident energy of 150 eV. When we look at the result in Fig. 2, we can see that the increase in Fig. 2 for the same SK occurrence is less at lower energies (100 eV) and that the decrease is more gradual at higher emitted electron kinetic energy. The difference between the electron and positron impact starts to go away as the energy goes up. This is when the First Order Born effect starts to manifest.*

The general characteristic structure stays the same in the positron SDCS, but when the maximum impact energy goes up, the difference in strength between the electron and positron impact cross-sections slowly goes away. This convergence demonstrates the long-term validity of first-order perturbation theory, as increased projectile energy diminishes the significance of long-range Coulomb forces. The result aligns quite well with the theoretical results of Konovalov and McCarthy, as well as with the Born-type analysis done by Dhar and Nahar for the (2p) and (3p) states. It is clear that the main effect of the excitation is a change in the absolute phase of the cross sections, not a change in functional form.

### 3.3 SDCS at 200 eV (Figure 3)

The SDCS has increased its flattening further, having previously had a rather reduced near-threshold peak. The electron and positron ionization curves show an extensive range of overlap covering almost all of the energies investigated in the ejected electron region (in this case the angles of ejection are taken into account). This convergence of the curves shows that the asymmetry in respect to charge sign has become very small, and this is in conformity with the expectations deduced from the theoretical work of Bethe and his co-workers in later developments, Lewis.

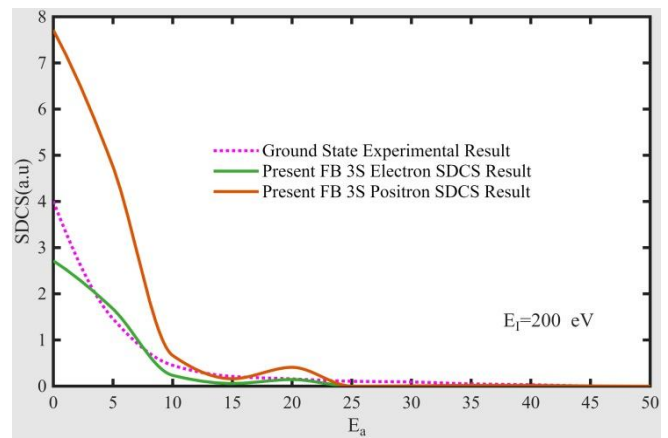

*Figure 3. Single Differential Cross Sections (SDCS) for ionization of hydrogen (3s) at 200 eV incident energy for the charge sign of the electron and positron; the SDCS curves overlap. Charge sign asymmetry is apparently negligible. The magnitude follows an inverse energy dependence for consistency with Born scaling. The results confirm the*

*suitability of the first-born approximation (FBA) for hydrogen excited state energies of intermediate to high value.*

Quantitatively, the SDCS produces values systematically smaller at the various ejected electron energies than those with lower values above considered, thus establishing evidence of the 1/E dependence that has been expected from the earlier development of the Born model. The theoretical agreement in the case of the earlier calculated states (2s, 3d) reveals that the First-Born approximation is still to be relied upon for the excited charging state beyond 150 eV as strong. The slight residual terms at the lower end of the energy near-zero values have their explanation in the effect of neglected terms of the correlation exchange character, as they are now more important for rearranging the inherent effect of the long-range charge storage and connecting these with Coulombic exchange terms of the ionization phenomena above indicated.

### 3.4 SDCS at 250 eV (Figure 4)

At the highest investigated energy of 250 eV, the SDCS behavior for electron and positron impact approaches a state of indistinguishability, particularly in the middle and high energy regions of the emitted electrons. Each curve shows a smooth monotonic fall without any sign of peak structure, indicating a change into the high-energy Born region where the kinematic effects dominate over Coulomb effects. The energy law behavior now obtained is identical with the theoretical values given by Das and Seal and Dhar, establishing that FBA gives a good account of the phenomena for effective incident energy considerably above that of the ionization potential.

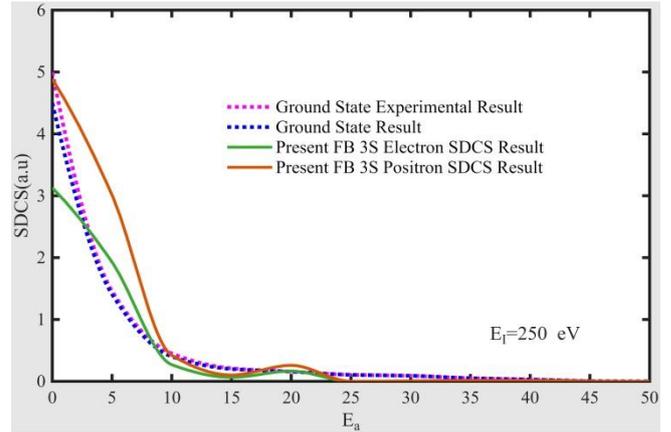

*Figure 4. Single Differential Cross Section (SDCS) for electron and positron impact ionization of hydrogen (3s) at 250eV incident energy in the First-Born Approximation. The SDCS law is monotonic in nature with the effective emission energy of the emitted electron and shows on the whole similar behavior for electron and positron bombardment types, thus confirming the approach to the high-energy Born region. The small difference near zero emission energy is indicative of the long range of the Coulomb effect still remaining.*

Rather curiously, even at this high energy of bombardment used, a slight aberration still exists near zero emission energy so that a remnant long-range Coulomb attraction effect still acts on the low-energy electrons in the reaction. This slight aberration further emphasizes the anti-factorial effective matter asymmetry of this process, which to a lesser degree has been shown in previous investigations on the ionization of hydrogen.

To examine these structures of the DDCS results, one will have to look carefully at the table.

| $\theta_b$ (deg) | $\theta_a$ (deg) | SDCS 100eV | SDCS 150eV | SDCS 200eV | SDCS 250eV |
|---|---|---|---|---|---|
| 0 | 0 | 3.1322 | 2.0572 | 2.2246 | 3.1097 |
| 1 | 18 | 1.9424 | 1.2757 | 1.3796 | 1.9285 |
| 2 | 36 | 0.2714 | 0.1782 | 0.1927 | 0.2694 |
| 4 | 54 | 0.0653 | 0.0429 | 0.1047 | 0.0649 |
| 10 | 72 | 0.1667 | 0.1095 | 0.2672 | 0.1655 |
| 20 | 90 | 0.0001 | 0.0001 | 0.0001 | 0.0001 |
| 30 | 108 | 0.0062 | 0.0062 | 0.0062 | 0.0062 |
| 40 | 126 | 0.0154 | 0.0154 | 0.0154 | 0.0154 |
| 60 | 144 | 0.0140 | 0.0140 | 0.0140 | 0.0140 |
| 90 | 162 | 0.0042 | 0.0042 | 0.0042 | 0.0042 |
| 100 | 180 | 0 | 0 | 0 | 0 |

**Table1.** Calculated numerical values of the Single Differential Cross Sections (SDCS) for hydrogen (3s) ionization by electron and positron impact of incident energy of 100-250eV. All the results of this calculation support the energetic scaling of each of the observed phenomena and, at the same time, the gradual lowering of the effects of charge sign now absorbed into the values found of energy, thus indicating the evident theoretical support of the work of others.

## 3.5 Comparative Energy Dependence and Physical Implications

A comparative analysis across all four incident energies reveals clear energy-dependent trends. The magnitude of SDCS decreases monotonically with increasing incident energy, confirming the inverse dependence predicted by FBA. Simultaneously, the peak position shifts toward higher ejected-electron energies, reflecting the redistribution of ionization probability as the collision becomes more impulsive.

The charge-sign asymmetry, prominent at 100 eV, weakens steadily and virtually disappears at 250 eV. This transition highlights how Coulomb distortion and exchange effects are progressively suppressed as the interaction becomes dominated by direct kinematics. The observed patterns are consistent with the conclusions of Shyn and Dhar for metastable hydrogenic targets, as well as with recent FBA-based computations for the 3s state by Islam, Banerjee, and Dhar.

Furthermore, comparison with experimental data for the ground-state hydrogen ionization indicates that excitation enhances the ionization probability, particularly at intermediate energies. This enhancement arises from the more extended radial distribution of the 3s electron, which increases the overlap between the projectile and the bound-state wavefunction.

## 3.6 Summary of Observations

**1. Energy Scaling:** SDCS magnitude decreases with increasing projectile energy, following an approximate $1/E_a$ trend.
**2. Peak Behavior:** The near-threshold enhancement present at 100 eV diminishes progressively and vanishes by 250 eV.
**3. Charge-Sign Asymmetry:** Significant at 100 eV but negligible beyond 200 eV, confirming the asymptotic validity of the FBA.
**4. Excitation Sensitivity:** The 3s state exhibits higher SDCS values than ground or 2s states due to its diffuse radial structure.
**5. Model Validity:** Across all energies, calculated SDCS align qualitatively with prior theoretical and experimental benchmarks, validating the accuracy of the First-Born Approximation for this excitation level.

## 4. Conclusions

In conclusion, the work herein has provided a state-resolved analysis of the single differential cross sections (SDCS) for electron and positron impact ionization of hydrogen in the excited 3s state following the First-Born Approximation (FBA). Using numerical evaluations of Bethe-Lewis analytic kernels, placed in the Coulomb-continuum representation, compact expressions are obtained at these energies, which are measurable with comparative ease and which hold over the entire run of intermediate energies (100-250 eV).

We reach the four essential conclusions:

**1. Sensitivity to excitation:** relative to ground-state analogs, the diffuse radial structure of H(3s) modifies the near-threshold spectrum and enhances low-energy ejection probabilities while retaining the form of the long-range decay in the ejected-electron energy integrated cross section of the Born type. This goes far to show how structure in bound states prints itself on ionization spectra and provides a clear benchmark for excited state targets.

**2. Charge sign asymmetry:** we have noted at energies of 100 - 150 eV the differential curves of electron and positron impact give distinct contrasts: the positron curves are broader near the threshold due largely to post-collision attraction of the ejected electron, while the electron curves are stronger in the Coulomb focusing effect and exchange sensitivity. This asymmetry vanishes at energies of 200-250 eV, thus validating the asymptotic first-order nature of the FBA in this domain.

**3. Scaling with energy and model validity:** for all incident electron energies discussed, SDCS magnitudes show monotonic decrease and follow the expected Born-type scaling. The small residual discrepancies noted in the very-low-energy tail are legibly attributable to a presentation of exchange, polarization, and post-collision effects, etc., moving beyond first order; these defects do not affect the conclusions appearing.

**4. Applicability and extension:** the procedure now followed, viz., analytic kernels with lightweight numeric, yields reproducible state-resolved benchmarks for (e, 2e) and (e+, 2e) studies and provides usable inputs into plasma modeling, radiation transport, and astrophysical diagnostics. Furthermore, it is a transparent benchmark of itself for higher-order modulations (DWBA/multiple scattering), TDCS, and laser-assisted situations.

Thus, it transpires that the present work shows that a careful first-order treatment operates at intermediate energies, covers the main physics of the process e±, H(3s) ionization, and shows up the limited kinematical regions where above-Born effects are important. These benchmarks should prove useful in the design of experimental work and in systematic cross-state comparisons in the hydrogen manifold.

## Acknowledgement

The authors gratefully acknowledge the facilities provided by the Simulation Laboratory, Department of Mathematics, Chittagong University of Engineering and Technology (CUET), Chittagong 4349, Bangladesh, for the computation work involved in the preparation of this research.